\documentclass[11pt]{article}
\usepackage{graphicx}
\begin{document}

\begin{center}
{\Large\bf Path integration in the field of a topological defect: the
case of dispiration}\\
~\\

Akira Inomata\\

Department of Physics, State University of New York at Albany\\
Albany, NY 12222, USA\\[2mm]

Georg Junker\\

European Organization for Astronomical Research in the Southern
Hemisphere\\
Karl-Schwarzschild-Strasse 2, D-85748 Garching, Germany\\[2mm]

James Raynolds\\

College of Nanoscale Science and Engineering\\
State University of New York at Albany\\
Albany, NY 12203, USA
\end{center}
~\\

\small{The motion of a particle in the field of dispiration (due to a
wedge disclination and a screw dislocation) is studied by path
integration. By formulating a $SO(2) \times T(1)$ gauge theory, first, we derive the
metric, curvature, and torsion of the medium of dispiration. Then we
carry out explicitly path integration for the propagator of a particle
moving in the non-Euclidean medium under the influence of a scalar
potential and a vector potential. We obtain also the winding number
representation of the propagator by taking the non-trivial topological
structure of the medium into account. We extract the energy spectrum and
the eigenfunctions from the propagator. Finally we make some remarks for
special cases. Particularly, paying attention to the difference between
the result of the path integration and the solution of Schr\"odinger's
equation in the case of disclination, we suggest that the Schr\"odinger
equation may have to be modified by a curvature term.}\\


Keywords - Path integral\\

\section{Introduction}

In recent years quantum effects on particle propagation in a field of
topological defects have attracted considerable attention (see, e.g.,
\cite{RS,BST,A}; see also a recent review article \cite{KF}).
Although the notion of ``defects'' in physics was originally associated
with crystalline irregularities, it has been extended to more general
topological structures such as entangled polymers, liquid crystals,
vortices, anyons, global monopoles, cosmic strings, domain walls, etc.
In 1950's, from the structuralogical aspect, Kondo {\it et
al}.\cite{Kon1} extensively studied unified geometrical treatments of
various subjects including elastic and plastic media, relativity,
network systems, and information theory. In particular, Kondo
\cite{Kon2} related dislocations to Cartan's torsion in the medium.
Since then the relation between dislocation theory and non-Riemannian
geometry has been well established. More recent approaches to defect
problems are gauge theories similar to those in particle physics
\cite{Kro,KE,Kl1,PS}. In 1978, Kawamura \cite{Kawa} pointed out that a
screw dislocation in a crystal produces an Aharonov-Bohm type effect in
particle scattering. The Aharonov-Bohm effect \cite{AB} is usually
understood as a topological effect \cite{SchT,Schbook,IS,GI}. Since
standard approaches to particle-defect interaction problems constitute of
solving relevant (local) differential equations, the role of topology
becomes often obscure.

In the present paper, we analyze the quantum behavior of a
particle in the vicinity of a topological defect by the path integral
method. Specifically we carry out path integration for a particle moving
around a dispiration\footnote{The Volterra process of forming a
dislocation and a disclination involves a translation and a rotation,
respectively. A {\it dispiration} is a defect formed by a translation and a
rotation at the same time \cite{Har}. See Fig.\ 1 and Fig.\ 2.}.  The dispiration we consider is a
composite structure of a screw dislocation and a wedge disclination with
a common defect line. In Sec.\ 2, we study the geometrical and topological
properties of the dispiration by using the $SO(2) \times T(1)$ gauge
theory. For the medium ${\cal M}$ of the dispiration, we derive the
squared line element,
\begin{equation}
ds^{2}=dr^{2} + \sigma ^{2}r^{2}d\theta ^{2} + (dz + \beta d\theta
)^{2}, \label{metric}
\end{equation}
where $\beta $ and $\sigma $ are the parameters directly related to the
Burgers vector of dislocation and the Frank vector of disclination,
respectively. The medium has a non-Euclidean structure with singular
torsion and curvature along the dispiration line.

In Sec.\ 3, we calculate in detail the path integral to obtain the
propagator (Feynman kernel) for a particle in the field of the
dispiration characterized by the line element (\ref{metric}) under the
influence of a scalar potential and a vector potential. From the
propagator so obtained, we extract the energy spectrum and the energy
eigenfunctions. In Sec.\ 4, converting the propagator in the partial wave
expansion into that in the winding number representation, we show that
the effect due to the multiply connected structure of the medium is
taken into account. The energy eigenfunctions and the energy spectrum are
extracted from the propagator in Sec.\ 5.
Sec.6 is devoted to interpretation of the energy
spectrum. In particular, we discuss the difference between the results
from the Schr\"odinger equation with no curvature term and those from
the path integral for the case of the conical space. Our path integral
calculation suggests that the standard Schr\"odinger equation may have
to be modified by a curvature term in order for two approaches to be
consistent.

\section{Gauge formulation of dispiration}

The dispiration under consideration is a combined structure of a screw
dislocation along the $z$-axis and a wedge disclination about the same
$z$-axis. See Figs. 1-2. The gauge theoretical approach to dislocation
and disclination has been extensively discussed in the literature
\cite{Kro,KE,Kl1,PS}. Here we wish to present a $SO(2) \times T(1)$
gauge approach to the dispiration developed along the line similar to
the formulation by Puntigam and Soleng \cite{PS}.\\

\noindent{\bf The $SO(2) \times T(1)$ gauge transformation:--} In the
gauge theoretical treatment, a deformation of an elastic medium in three
dimensions is described by a local coordinate transformation consisting
of a rotation and a translation,
\begin{equation}
{\bf x}'=\mbox{\boldmath $\rho $}({\bf x})\cdot{\bf x} + \mbox{\boldmath
$\tau $}({\bf x}), \label{deform}
\end{equation}
where $\mbox{\boldmath $\rho  $}\in SO(2)$, $\mbox{\boldmath ${\cal
\tau}$}\in T(1)$, and ${\bf x}=\{ x, y, z\}$ is the position vector
of the undeformed three-dimensional medium.

As is well-known, an axial wedge disclination can be created by the
so-called Volterra process: That is, by (a) removing a wedge shaped portion
of an angle $\gamma \in [0, 2\pi )$ from the medium and pasting the open
walls together, or (b) inserting an extra wedge-shaped portion with
$\gamma \in [-2\pi, 0)$ into the medium. Note that the deficit angle
$\gamma $ is chosen to be positive for the case (a) and negative for the
case (b). See Fig. 1 for the case (a). The wedge disclination about the
$z$-axis is a rotational deformation obtained by gauging $SO(2)$:
\begin{equation}
{\bf x}'=\mbox{\boldmath $\rho  $}(\theta)\cdot{\bf x} \label{rot}
\end{equation}
with the rotation matrix,
\begin{equation}
\mbox{\boldmath $\rho  $}(\theta )=
\left(\begin{array}{ccc}~\cos (\gamma \theta /2\pi ) &
-\sin (\gamma \theta /2\pi ) & 0 \\
\sin (\gamma \theta /2\pi ) & \cos (\gamma \theta /2\pi ) & 0 \\
0 & 0 & 1 \end{array}\right). \label{rho}
\end{equation}
Here $\theta = \tan^{-1}(y/x) \in [0, 2\pi )$.

A screw dislocation lying along the $z$-axis (i.e., having a constant
Burgers vector pointing the $z$-direction) as shown in Fig. 2 is a
translational deformation obtained by gauging the $z$-translational
group $T(1)$:
\begin{equation}
{\bf x}'={\bf x} + \mbox{\boldmath $\tau $}(\theta )
\end{equation}
with the angle-dependent $z$-translation vector
\begin{equation}
\mbox{\boldmath $\tau $}(\theta )=-\frac{b\theta }{2\pi }{\bf e}_{z}
\label{trans}
\end{equation}
where $b$ is a translation parameter and $\theta =\tan^{-1}(y/x)$ as
before.

The dispiration comprised of such a wedge disclination and a screw
dislocation is described by the combined coordinate transformation,
\begin{equation}
{\bf x}'=\mbox{\boldmath $\rho $}(\theta )\cdot{\bf x} + \mbox{\boldmath
$\tau $}(\theta )   \label{com}
\end{equation}
where $\mbox{\boldmath $\rho $}(\theta )$ and $\mbox{\boldmath
$\tau $}(\theta )$ are given by (\ref{rho}) and (\ref{trans}),
respectively.\\

\noindent {\bf Gauge connections:--}
The standard Yang-Mills theory localizes with respect to the external
space the gauge group which acts homogeneously in the internal space and
introduces the gauge potential or the connection ${\bf\Gamma} $ that
transforms under the group action ${\cal G}$ as
\[
{\bf \Gamma }' = {\cal G}{\bf \Gamma }{\cal G}^{-1} - d{\cal G}\,{\cal
G}^{-1}.
\]
What we wish to formulate for the dispiration is a $SO(2)\times T(1)$
gauge approach which differs in character from the Yang-Mills theory; the
gauge group acts inhomogeneously on the (internal) coordinates ${\bf x}$
that is soldered locally to the (external) coordinates of the medium of
dispiration. It is more appropriate for us to follow the
procedures used in constructing a Poincar\'e gauge theory \cite{PS,T82} or an
affine gauge theory \cite{H95} for gravity.

First we write the group action (\ref{com}) in the matrix form,
\[
{\cal G}\bar{\bf x}=
\left(\begin{array}{cc}
{\mbox{\boldmath $\rho$}} & {\mbox{\boldmath $\tau$}} \\
0 & 1 \end{array}
\right)
\left(\begin{array}{c}{\bf x}\\ 1 \end{array} \right) =
\left(\begin{array}{c}
{\mbox{\boldmath $\rho$}}\cdot {\bf x} + {\mbox{\boldmath $\tau$}} \\ 1 \end{array}
\right).
\]
Then we define the connection ${\bf \Gamma }$ as
\[
{\bf\Gamma} =\left(\begin{array}{cc} {\bf\Gamma}^{(R)} & {\bf\Gamma} ^{(T)} \\ 0 & 0
\end{array} \right)
\]
where ${\bf\Gamma} ^{(R)}$ and ${\bf\Gamma}^{(T)}$ are the rotational connection
and the translational connection, respectively. In order for ${\bf\Gamma} $ to
behave like a connection in the Yang-Mills theory, ${\bf\Gamma} ^{(R)}$ and
${\bf\Gamma}^{(T)}$ must transform as
\[
{\bf\Gamma} ^{(R)\prime}={\mbox{\boldmath $\rho$}}{\bf\Gamma} ^{(R)}{\mbox{\boldmath $\rho$}}^{-1} -
(d{\mbox{\boldmath $\rho$}}){\mbox{\boldmath $\rho$}}^{-1}
\]
and
\[
{\bf\Gamma}^{(T)\prime}={\mbox{\boldmath $\rho$}}{\bf\Gamma}^{(R)} - d{\mbox{\boldmath $\tau$}}-
\left[{\mbox{\boldmath $\rho$}} {\bf\Gamma} ^{(R)} {\mbox{\boldmath $\rho$}}^{-1} -
(d{\mbox{\boldmath $\rho$}})\,{\mbox{\boldmath $\rho$}}^{-1}\right]{\mbox{\boldmath $\tau$}} .
\]
Evidently the rotational connection is a $SO(2)$-valued one-form, and
the translational connection is a ${\bf R}$-valued connection one-form.

Now we construct the solder form that locally connects the global gauge coordinates
${\bf x}(0)$ to the coordinates ${\bf x(\theta )}$ of the dispiration
medium as
\begin{equation}
\mbox{\boldmath $\omega $} = d{\bf x} + \mbox{\boldmath $\Gamma
$}^{(R)}\cdot{\bf x} + \mbox{\boldmath $\Gamma $}^{(T)} \label{omega}
\end{equation}
which is a vector valued one-form.\footnote{It is possible to derive
this solder form by starting with the homogeneous group $SO(4)$ and by
applying the group contraction to reduce it to $SO(3)\times T(1)$. The
solder form may be obtained as a limiting form of a component of the
connection. For instance, the de Sitter gauge theory can be
contracted to the Poincar\'e gauge theory, so that the solder form
is obtained from the gauge potential as the vierbeine \cite{I79}. This
contraction scheme does not work for the affine gauge theory \cite{H95}.}
It transforms as $\mbox{\boldmath $\omega$}' = \mbox{\boldmath $\rho$}
\mbox{\boldmath $\omega$}$, leaving $g={\mbox{\boldmath $\omega$}}^{T}\cdot {\mbox{\boldmath $\omega $}}$ invariant. Since we
create the dispiration by a gauge transformation in flat space, we
choose a connection that vanishes at $\theta =0$. Then the corresponding
rotation and translation connections are given, respectively, by
\begin{equation}
\mbox{\boldmath $\Gamma $}^{(R)} =\mbox{\boldmath $\rho $}
\,d\mbox{\boldmath $\rho $}^{-1}\label{rotconn}
\end{equation}
and
\begin{equation}
\mbox{\boldmath $\Gamma $}^{(T)} = -d\mbox{\boldmath $\tau $}.\label{transconn}
\end{equation}
The rotational connection (\ref{rotconn}) and the translational connection (\ref{transconn}) can
be easily calculated by using the rotation matrix (\ref{rho}) and the
translation vector (\ref{trans}). The differential of the rotation matrix (\ref{rho}) is
\begin{equation}
d\mbox{\boldmath $\rho  $}^{-1}(\theta )=
d\mbox{\boldmath $\rho  $}(-\theta )= -\frac{\gamma }{2\pi }\,d\theta
\,
\left(\begin{array}{ccc}~ \sin (\gamma \theta /2\pi ) &
-\cos (\gamma \theta /2\pi ) & 0 \\
\cos (\gamma \theta /2\pi ) & \sin (\gamma \theta /2\pi ) & 0 \\
0 & 0 & 0 \end{array}\right)
\end{equation}
where
\begin{equation}
d\theta = \frac{1}{r^{2}}(x\,dy - y \,dx) ~,~~~r^{2} = x^{2} + y^{2}.
\label{dtheta}
\end{equation}
This and the rotation matrix (\ref{rho}) together lead us to
the rotational connection,
\begin{equation}
\mbox{\boldmath $\Gamma $}^{(R)} =\mbox{\boldmath $\rho  $}(\theta )\,
d\mbox{\boldmath $\rho  $}^{-1}(\theta ) =\frac{\gamma }{2\pi }\,{\bf
m}\,d\theta \label{rconn}
\end{equation}
where
\begin{equation}
{\bf m}=\left(\begin{array}{rcc}~ 0 & ~1 & ~0~ \\
~-1 & ~0 & ~0~ \\
~0 & ~0 & ~0~\end{array}\right).
\end{equation}
The translational connection is found in a simple form by
differentiating the translation vector (\ref{trans}),
\begin{equation}
\mbox{\boldmath $\Gamma $}^{(T)}= -d\mbox{\boldmath $\tau $} =
\frac{b}{2\pi }\,d\theta \,{\bf e}_{z}. \label{tconn}
\end{equation}
Substitution of (\ref{rconn}) and (\ref{tconn}) into (\ref{omega})
yields
\begin{equation}
\mbox{\boldmath $\omega $}=d{\bf x} + \frac{\gamma }{2\pi }\,
d\theta \,{\bf m}\cdot {\bf x} + \frac{b}{2\pi }\,d\theta \,{\bf e}_{z}
= \left(\begin{array}{lll}dx &+ & (\gamma /2\pi )y\,d\theta \\
dy &- &(\gamma /2\pi )x\, d\theta \\
dz &+ & ~ (b/2\pi )\, d\theta \end{array}\right). \label{omega2}
\end{equation}
Utilizing this solder form as the coframe we can determine the
squared line element $ds^{2}=g_{ij}\,dx^{i}\otimes dx^{j}$ in the medium
of the dispiration,
\begin{equation}
ds^{2} = \delta _{\alpha \beta }\omega ^{\alpha }\otimes \omega ^{\beta }
=dr^{2} + \sigma ^{2}\,r^{2}\,d\theta ^{2} +
(dz + \beta d\theta )^{2}, \label{line2}
\end{equation}
where
\begin{equation}
\sigma = 1 - \frac{\gamma }{2\pi }, ~~~~~\beta = \frac{b}{2\pi }.
\label{si-be}
\end{equation}
Evidently $\sigma \in (0, 1]$ when $\gamma \in [0, 2\pi )$, and $\sigma
\in (1, 2]$ when $\gamma \in [-2\pi , 0)$.

In the present paper, we consider the non-simply connected medium ${\cal M}={\bf R}^3\backslash \{x=y=0\}$ with metric (\ref{line2}) as the field of dispiration. In the following, we briefly review some of the geometrical and topological properties that will be useful for later discussion.\\

\noindent {\bf Curvature and Frank vector:--} The curvature two-form ${\bf R}$ is defined in terms of the rotational connection $\mbox{\boldmath $\Gamma $}^{(R)}$,
\begin{equation}
{\bf R}=d\mbox{\boldmath $\Gamma $}^{(R)} + \mbox{\boldmath $\Gamma $}^{(R)}
\wedge \mbox{\boldmath $\Gamma $}^{(R)}.  \label{curv}
\end{equation}
For the case of dispiration with (\ref{rconn}), since $ \mbox{\boldmath
$\Gamma $}^{(R)} \wedge \mbox{\boldmath $\Gamma $}^{(R)}=0$,
it integrates into
\begin{equation}
\int_S{\bf R}=\int_Sd\mbox{\boldmath $\Gamma $}^{(R)}=\int_{\partial S} \mbox{\boldmath $\Gamma $}^{(R)}=
\int_{\partial S}\frac{\gamma }{2\pi }\,{\bf m}\,d\theta=\gamma{\bf m} ,
\label{intcurv}
\end{equation}
where $\partial S$ denotes the boundary of a surface $S$. Now we choose an orthogonal frame $(\xi^1,\xi^2,\xi^3)$ such that $ds^2=\delta_{kl}\xi^k\wedge\xi^l$ with $\xi^1=dr$, $\xi^2=\sigma rd\theta$, $\xi^3=dz +\beta d\theta$. Then we have ${\bf R}={\bf R}_{kl}\xi^k\wedge\xi^l$. The surface is chosen to be orthogonal to $\xi^3$. Eq. (\ref{intcurv}) implies that
\begin{equation}
{\bf R}= \gamma \,{\bf m}\,\delta ^{(2)}(\xi^1,\xi^2)\,d\xi^1 \wedge d\xi^2 . \end{equation}
The corresponding scalar curvature is
\begin{equation}
R=2\gamma \delta ^{(2)}(\xi^1,\xi^2). \label{s-curve}
\end{equation}
or equivalently
\begin{equation}
R=2\frac{\gamma}{\sigma} \delta ^{(2)}(x,y). \label{s-curve2}
\end{equation}
Apparently the curvature of the medium ${\cal M}$ is zero everywhere
except along the $z$-axis ($x=y=0$). Note that the curvature is created
not by the dislocation but by the disclination.

The Frank vector ${\bf f}=\{\Phi ^{23}, \Phi ^{31}, \Phi ^{12}\}$
is defined to characterize a disclination with
\begin{equation}
\mbox{\boldmath $\Phi $} = \int _{_{S}} {\bf R}=\gamma \,{\bf m}
\end{equation}
where again the integral is over a surface $S$ which is delimited by a
loop $\partial S$ enclosing the $z$-axis. Obviously the only non-vanishing components of the Frank vector for the dispiration is given by $\Phi ^{12}$.
As a result, the Frank vector takes the form,
\begin{equation}
{\bf f}=\gamma \,{\bf e}_{z},
\end{equation}
which, points the $z$-direction and its magnitude
is identical to the deficit angle $\gamma $.

\noindent{\bf Torsion and Burgers vector:--}
The torsion two-form is defined by
\begin{equation}
{\bf T}=d\mbox{\boldmath $\omega $} + \mbox{\boldmath $\Gamma $}^{(R)}
\wedge \mbox{\boldmath $\omega $} = {\bf R}\cdot{\bf x} +
d\mbox{\boldmath $\Gamma $}^{(T)} + \mbox{\boldmath $\Gamma $}^{(R)}
\wedge \mbox{\boldmath $\Gamma $}^{(T)}, \label{tors}
\end{equation}
which, in general, depends not only on the translational connection but also on the
rotational connection. For (\ref{rconn}) and (\ref{tconn}), $
\mbox{\boldmath $\Gamma $}^{(R)} \wedge \mbox{\boldmath $\Gamma
$}^{(T)}=0$. Note also that the integration of ${\bf R}\cdot{\bf x}$ vanishes.
Then the torsion two-form integrates into
\begin{equation}
\int_{_{S}}{\bf T}
=\int_{_{S}}d{\bf \Gamma}^{(T)}
= \int_{\partial S}{\bf \Gamma}^{(T)}
=b{\bf e}_z.
\label{inttors}
\end{equation}
from which follows
\begin{equation}
{\bf T}= b\,{\bf e}_{z}
\delta^{(2)}(\xi^1,\xi^2)d\xi^1\wedge d\xi^2 . \label{tors3}
\end{equation}
We see that the only non-vanishing component of the torsion two-form is in the $z$-direction and is contributed only by the screw dislocation.

The Burgers vector ${\bf b}$ is defined by the surface integral of the torsion two-form
\begin{equation}
{\bf b}=\int_{_{S}}{\bf T}
\label{burg}
\end{equation}
which has been evaluated in (\ref{inttors})
\begin{equation}
{\bf b}=b{\bf e}_z.
\label{burg2}
\end{equation}
As is expected the translation parameter $b$ of (\ref{trans}) is
indeed the magnitude of the Burgers vector ${\bf b}$.\\

\noindent{\bf Conical space:--} Before closing this section, we consider
a special case of the line element (\ref{metric}) with constant $z$ and $\beta =0$. The two
dimensional surface having the metric
\begin{equation}
dl^{2}=dr^{2} + \sigma ^{2}r^{2}d\theta ^{2} \label{metl}
\end{equation}
to which (\ref{metric}) reduces may be realized as a conical
surface ${\sl M}_{c}$ if the surface is imbedded into a three
dimensional Euclidean space $E^{3}$. Let
\begin{equation}
{\bf X}=\left(\sigma r \cos \theta , \sigma r \sin \theta, \sqrt{1-
\sigma ^{2}}\,r\right).
\end{equation}
Apparently,
\begin{equation}
I: ~d{\bf X}\cdot d{\bf X}=dl^{2}
\end{equation}
which is the first fundamental form of the imbedded surface $M_c$.
Again, parameterizing the surface by $0< r < \infty $ and $0 \leq \theta
< 2\pi $, we have the metric tensor and its inverse,
\begin{equation}
g_{ab}=\left(\begin{array}{cc}1 &0 \nonumber \\ 0 & \sigma
^{2}r^{2}\end{array}\right), ~~~~~~~~
g^{ab}=\left(\begin{array}{cc}1 &0 \nonumber \\ 0 & \sigma
^{-2}r^{-2}\end{array}\right),
\end{equation}
which will be used later. With
\begin{equation}
{\bf X}_{r}=\left(\sigma \cos \theta , \sigma \sin \theta, \sqrt{1-
\sigma ^{2}}\right),
\end{equation}
\begin{equation}
{\bf X}_{\theta }=\left(-\sigma r \sin \theta , \sigma r \cos \theta, 0
\right),
\end{equation}
the unit vector normal to ${\sl M}_{c}$ at a point $(r, \theta )$,
\begin{equation}
{\bf n}={\bf X}_{r}\times {\bf X}_{\theta }/|{\bf X}_{r}\times {\bf
X}_{\theta }|
\end{equation}
is easily calculated to be
\begin{equation}
{\bf n}=\left(-\sqrt{1-\sigma ^{2}}\cos \theta , -\sqrt{1-\sigma
^{2}}\sin \theta , \sigma\right).
\end{equation}
The second fundamental form,
\begin{equation}
II: ~-d{\bf X}\cdot d{\bf n}=G_{ab}du^{a} \otimes du^{b},
\end{equation}
is also immediately obtained for the conical surface ${\sl M}_{c}$ as
\begin{equation}
-d{\bf X}\cdot d{\bf n} = \sqrt{1 - \sigma ^{2}}\,\sigma r\,d\theta^{2}.
\end{equation}
Hence
\begin{equation}
{\bf G}=\left(\sum_{b}g^{ab}G_{bc}\right)=\left(\begin{array}{cc}0 &0
\nonumber \\ 0 & \sqrt{1-\sigma ^{2}}/(\sigma r )\end{array}\right)
\end{equation}
whose two eigenvalues $k_{1}=0$ and $k_{2}=\sqrt{1-\sigma ^{2}}/(\sigma
r)$ are the principal curvatures of the conical surface for $r \neq 0$.
The Gaussian curvature and the mean curvature of a two dimensional
surface imbedded in $E^{3}$ are defined, respectively, by
\begin{equation}
K= det {\bf G} = k_{1}k_{2}
\end{equation}
\begin{equation}
H=\frac{1}{2} tr {\bf G} = \frac{1}{2}(k_{1} + k_{2}).
\end{equation}
For the conical surface in question,
\begin{equation}
K=0 ~~~~~\mbox{and} ~~~~~H=\sqrt{1-\sigma ^{2}}/(2\sigma r)
\label{K}
\end{equation}
for $r \neq 0$. Of course the Gaussian curvature $K$ does not vanish at
the apex of the cone. According to the Gauss-Bonnet theorem,
\begin{equation}
\int_{S}K da + \int_{\partial S}k_{g} dl =
2\pi \chi (S)
\end{equation}
where $S$ is a compact two-dimensional Riemann manifold with
boundary $\partial S$, $K$ is the Gaussian curvature of
$S$, $k_{g}$ is the geodesic curvature of $\partial S$,
and $\chi (S)$ is the Euler characteristic of $S$.
Now we let $S$ be the lateral surface of a frustum with slant height $r$.
Then $k_{g} = 1/r$ and the line element $dl$ integrated along the
boundary $\partial S$ with $dr=0$ result in
\begin{equation}
\int_{\partial S}k_{g} dl=2\pi \sigma=2\pi - \gamma
\end{equation}
where $\gamma $ is the deficit angle defined in (\ref{si-be}). The Euler characteristic
of a cone is unity. Hence we have
\begin{equation}
\int_{S}K da = \gamma
\end{equation}
where $da=\sigma rdr\,d\theta $.
From this and (\ref{K}) follows
\begin{equation}
K = \frac{\gamma}{\sigma}\, \delta ^{2}(x, y)
\label{GaussCurv}
\end{equation}
which includes the curvature at the apex $(r=0)$.
As is well-known, the Ricci or scalar curvature of a two-dimensional manifold is twice the Gaussian curvature. Indeed, twice the Gaussian curvature (\ref{GaussCurv}) coincides with the scalar curvature (\ref{s-curve2}).

\section{Path integration}

The medium surrounding the dispiration considered in the
preceding section is ${\cal M}={\bf R}^3\backslash\{x=y=0\}$, which is geometrically a non-Riemannian space and
topologically a non-simply connected space. In this section, we carry out
path integration for the propagator of a particle moving in the field of
the dispiration. To confine the particle in the vicinity of the
dispiration, we introduce a two-dimensional harmonic oscillator
potential. Furthermore, we assume a repulsive inverse-square potential
to prevent the particle from falling into the singularity at the defect
line. For the purpose of comparison, we introduce also a vector
potential due to a flux tube and a uniform magnetic field.

The standard approach deals with the Schr\"odinger equation in curved space. As will be discussed in Sec.\ 5, the Schr\"odinger equation is usually modified in curved space not only by the Laplace-Beltrami operator replacing the Laplacian but also the so-called curvature term added as an effective potential. The energy spectrum is sensitive to the type of curvature term; yet the controversy on the choice of the term is not fully settled. The path integral calculation we present suggests that the Gaussian curvature of the surface where the particle moves would dominate the curvature term.

\subsection{The Lagrangian}

Now that the dispiration field is characterized by the line element
(\ref{line2}), the Lagrangian for a charged point particle of mass $M$
moving in the vicinity of the dispiration under the influence of a
scalar potential $V({\bf x})$ and a vector potential ${\bf A}({\bf x})$
is written as
\begin{equation}
L=\frac{1}{2}M \left(\frac{ds}{dt}\right)^{2}  -
\frac{e}{c}\dot{\bf x}\cdot {\bf A}({\bf x}) - V({\bf x}). \label{Lag1}
\end{equation}
As has been mentioned before, we choose the vector potential ${\bf A}({\bf
x})$ consisting of two parts; one due to an ideally thin flux tube
that contains constant magnetic flux $\Phi $ along the dispiration line,
and another due to a uniform constant magnetic field ${\bf B}=B\,{\bf
e}_{z}$ pointing the $z$-direction. Then the vector potential term of
the Lagrangian (\ref{Lag1}) is expressed in cylindrical coordinates
$(r, \theta , z)$ as
\begin{equation}
\frac{e}{c}\dot{\bf x}\cdot {\bf A}=\alpha \hbar \dot{\theta } + M\omega
_{_{L}}r^{2}\dot{\theta}
\end{equation}
where
\begin{equation}
\alpha = \frac{e\Phi }{2\pi \hbar c}, ~~~~~~\omega
_{_{L}}=\frac{eB}{2Mc}.
\end{equation}
Here $\alpha $ is the ratio of the magnetic flux to the fundamental
fluxon $\Phi _{0}=2\pi \hbar c/e$, which is identical to the statistical
parameter in Wilczek's anyon model \cite{Wilc}, and $\omega _{_{L}}$ is
the Lamor frequency. The flux tube is included in our calculation in
order to observe the similarity between the Aharonov-Bohm effect and
the effect of a screw dislocation. In the scalar potential $V({\bf x})$,
we include a two-dimensional short range repulsive potential (the
inverse square potential with $\kappa>0$ sufficiently large) to emphasize the impenetrable feature of the
central singularity, and a two-dimensional long range attractive potential (the
harmonic oscillator potential) to confine the particle in the vicinity
of the dispiration. Namely, $V({\bf x})$ is specified to be a
two-dimensional central force potential,
\begin{equation}
V(r)=\frac{\kappa \hbar^{2}}{8M\sigma ^{2}r^{2}} +
\frac{1}{2} M\omega_{0}^{2}r^{2} ~,~~~~r^{2}=x^{2}+y^{2}.
\end{equation}
Thus the Lagrangian we consider is
\begin{equation}
L=\frac{1}{2}M\left\{\dot{r}^{2}+ \sigma ^{2}r^{2}\dot{\theta }^{2}
+\left(\dot{z}+\beta \dot{\theta }\right) ^{2}\right\} - \alpha \hbar
\dot{\theta } - M\omega _{_{L}}\,r^{2}\dot{\theta } - V(r).
\label{Lag2}
\end{equation}

\subsection{The propagator}

The transition amplitude (propagator) for the three-dimensional motion
of the charged particle from point ${\bf x}^{\prime }= (r',\theta ',
z')$ to point ${\bf x}^{\prime \prime }=(r'', \theta '', z'')$ can be
calculated by the path integral \cite{Feyn}
\begin{equation}
K\left( \mathbf{x}^{\prime \prime }, \mathbf{x}^{\prime }; \tau \right)
=\int\nolimits_{{\bf x}^{\prime }={\bf x}(t^{\prime })}^{{\bf x}^
{\prime \prime }={\bf x}(t^{\prime
\prime })}\,
\exp \left[\frac{i}{\hbar }\int\nolimits_{t^{\prime}}^{t^{\prime \prime }}
\,L\,dt \right] {\cal D}^{3}{\bf x}
\end{equation}
where $\tau =t^{\prime \prime }-t^{\prime } > 0$. The integral measure
must be so chosen that the propagator satisfies the properties,
\begin{equation}
\lim_{t''\rightarrow t'}K\left( {\bf x}^{\prime \prime
},{\bf x}^{\prime };t''-t'\right)=\delta (\mathbf{x}^{\prime \prime
} - \mathbf{x}^{\prime }),
\end{equation}
\begin{equation}
\int K\left( {\bf x}^{\prime \prime
},{\bf x};t''-t\right)\,K\left( {\bf x}
,{\bf x}^{\prime };t-t^\prime\right)\,d^{3}{\bf x}
=K\left( {\bf x}^{\prime \prime
},{\bf x}^{\prime };t''-t'\right).
\end{equation}
The path integral we calculate with the Lagrangian (\ref{Lag2}) is
\begin{eqnarray}
\lefteqn{K\left({\bf r}'', z'' ; {\bf r}', z'; \tau \right)} \nonumber
\\
&&=\int \, \exp \left[ \frac{i}{\hbar }\int\nolimits_{t'}^{t''}\left\{
\frac{M}{2}\left(\dot{r}^{2} + \sigma ^{2}r^{2}\dot{\theta }^{2}\right)
+ \frac{M}{2}\left(\dot{z} + \beta \dot{\theta }\right)^{2} - \alpha
\hbar \dot {\theta } - M \omega _{_{L}}r^{2}\dot{\theta } -
V(r) \right\}
dt\right]\,{\cal D}^{2}{\bf r}\,{\cal D}z. ~~~~~~  \label{3prop}
\end{eqnarray}
In (\ref{3prop}) ${\bf r}$ signifies two variables $(r, \theta
)$ symbolically, and the two dimensional integral measure $d^{2}{\bf
r}$ will be specified later. After path integration, from the
propagator, we should be able to extract the energy spectrum and the
wave functions for the system.

\subsection{The $z$-integration}

First we perform the $z$-integration by letting $\zeta =z+\beta
\theta $. The $z$-path integral is nothing but the Gaussian path
integral for a one dimensional free particle, which yields the standard
result,
\begin{equation}
\int\nolimits_{\zeta ^{\prime }=\zeta \left( t^{\prime }\right) }^{\zeta
^{\prime \prime }=\zeta \left( t^{\prime \prime }\right) }\exp \left[
\frac{ i}{\hbar }\int\nolimits_{t^{\prime }}^{t^{\prime \prime }}
\frac{M}{2}\dot{ \zeta }^{2}dt\right] {\cal D}\zeta =\sqrt{\frac{M}{2\pi
i\hbar \tau }}\exp \left[ \frac{iM\left( \zeta ^{\prime \prime }-\zeta
^{\prime }\right) ^{2}}{2\hbar \tau }\right]. \label{z-int}
\end{equation}
Now we rewrite the right hand side of (\ref{z-int}) as
\begin{equation}
\sqrt{\frac{M}{2\pi i\hbar \tau }}\exp \left[ \frac{iM\left( \zeta ^{\prime
\prime }-\zeta ^{\prime }\right) ^{2}}{2\hbar \tau }\right] =\frac{1}{2\pi }%
\int\nolimits_{-\infty }^{\infty }e^{-i\tau\hbar k^{2}/2M}e^{i\left( \zeta
^{\prime \prime }-\zeta ^{\prime }\right) k}dk  \label{Edecomp}
\end{equation}%
where $\hbar k$ is the $z$-component of momentum of the particle. We also
notice that
\begin{equation}
\zeta ^{\prime \prime }-\zeta ^{\prime }=z^{\prime \prime }-z^{\prime
}+\beta ^{\prime }\left( \theta ^{\prime \prime }-\theta ^{\prime }\right)
=z^{\prime \prime }-z^{\prime }+\beta \int\nolimits_{t^{\prime
}}^{t^{\prime \prime }}\dot{\theta }dt .
\end{equation}
Incorporating these results into the path integral (\ref{3prop}), we
decompose the propagator as
\begin{equation}
K\left( \mathbf{x}^{\prime \prime },\mathbf{x}^{\prime };\tau \right)
=\frac{ 1}{2\pi }\int\nolimits_{-\infty }^{\infty }dke^{ik\left(
z^{\prime \prime }-z^{\prime }\right) }e^{-i\tau \hbar
k^{2}/2M}K^{\left( k\right) }\left( \mathbf{r}^{\prime \prime
},\mathbf{r}^{\prime };\tau \right) \label{3pro2}
\end{equation}
with the two-dimensional propagator for a fixed $k$ value,
\begin{eqnarray}
\lefteqn{K^{\left( k\right) }\left( {\bf r}^{\prime \prime }, {\bf r}^{\prime
};\tau \right)}\nonumber \\
&&=\int^{{\bf r}''={\bf r}(t'')}_{{\bf r}'={\bf r}(t')} \exp \left\{ \frac{i}{\hbar }
\int\nolimits_{t^{\prime }}^{t^{\prime \prime }}\left[ \frac{M}{2}
\left(\dot{r}^{2} + \sigma ^{2}r^{2}\dot{\theta }^{2}\right) - M\omega
_{_{L}}r^{2}\dot{\theta } - \xi \hbar
\dot{\theta } -V(r)\right] dt\right\} {\cal D}^{2}{\bf r}.~~ \label{2K}
\end{eqnarray}
where $\xi = \alpha - \beta k$. The integral on the right hand side of
(\ref{Edecomp}) is the spectral decomposition of the $z$-motion with the
continuous spectrum,
\begin{equation}
E_{k}= \frac{\hbar^{2}k^{2}}{2M}. \label{zspec}
\end{equation}
Next we make a change of the angular variable from $\theta $ to $\vartheta $
by letting
\begin{equation}
\dot{\vartheta}=\dot{\theta } - \bar{\omega}
\end{equation}
where $\bar{\omega }= \omega _{_{L}}/\sigma ^{2}$. Accordingly the
$k$-propagator (\ref{2K}) is transformed into
\begin{eqnarray}
\lefteqn{K^{\left( k\right) }\left( {\bf r}^{\prime \prime }, {\bf r}^{\prime
};\tau \right)}\nonumber \\
&&=\int^{{\bf r}''={\bf r}(t'')}_{{\bf r}'={\bf r}(t')} \exp \left\{
\frac{i}{\hbar } \int\nolimits_{t^{\prime }}^{t^{\prime \prime }}\left[
\frac{M}{2} \left(\dot{r}^{2} + \sigma ^{2}r^{2}\dot{\vartheta
}^{2}\right) - \xi \hbar \dot{\vartheta } -
U(r)\right] dt\right\}{\cal D}^{2}{\bf r}(r, \vartheta ) \label{2K2}
\end{eqnarray}
where
\begin{equation}
U(r)= V(r) + \frac{1}{2}M\bar{\omega }^{2}r^{2} - \xi \hbar
\bar{\omega},
\end{equation}
or
\begin{equation}
U(r) =\frac{\kappa \hbar^{2}}{8M\sigma ^{2}r^{2}} +
\frac{1}{2} M\omega^{2}r^{2} + V_{0}
\end{equation}
where
\begin{equation}
\omega ^{2} = \omega _{0}^{2} + \bar{\omega }^{2}, ~~~~~V_{0}=-\xi
\hbar \bar{\omega }.
\end{equation}
Note that $\vartheta $ is the angular variable in a rotating frame with
angular velocity $ -\bar{\omega}$. For simplicity the constant $V_{0}$ in the potential will be ignored in the calculation below.

\subsection{The short time action}

To calculate the two-dimensional path integral (\ref{2K2}) in polar
coordinates \cite{EG,PI,BJ}, we first express it
in discretized form,
\begin{equation}
K^{\left( k\right) }\left({\bf r}^{\prime \prime }, {\bf r}
^{\prime };\tau \right) =\lim_{N\rightarrow \infty }
\int\nolimits_{{\bf r}^{\prime
}={\bf r}(t^{\prime })}^{{\bf r}^{\prime \prime }={\bf r}(t^{\prime \prime })}
\prod\limits_{j=1}^{N}K^{(k)}\left( {\bf r}_{j},{\bf r}
_{j-1};\epsilon \right) \prod\limits_{j=1}^{N-1}d^{2}\mathbf{r}_{j}
\label{disK}
\end{equation}
where the propagator for a short time interval $\epsilon =t_{j}-t_{j-1}=\tau
/N $ is given by
\begin{equation}
K^{(k)}\left( {\bf r}_{j},{\bf r}_{j-1};\epsilon \right)
=A_{j}\exp \left( \frac{i}{\hbar }S_{j}\right). \label{spro}
\end{equation}
Now we select a relevant approximation of the short time action
$S_{j}$ in cylindrical coordinates, and determine the amplitude $A_{j}$
so as to meet the normalization condition
\begin{equation}
\lim_{\epsilon  \rightarrow 0}K^{(k)}\left(
{\bf r}_{j},{\bf r}_{j-1};\epsilon \right) = \delta ^{(2)}(
{\bf r}_{j} - {\bf r}_{j-1}) \label{norm2}
\end{equation}
where the two-dimensional delta function satisfies
\begin{equation}
\int \delta ^{(2)}({\bf r}_{j} - {\bf r}_{j-1})\,d^{2}{\bf r}_{j}=1.
\label{delta}
\end{equation}
The short time action is
\begin{equation}
S_{j}=\int\nolimits_{t_{j-1}}^{t_{j}}\left[ \frac{M}{2}(\dot{
r}^{2} + \sigma ^{2}r^{2}\dot{\vartheta }^{2}) - \xi \hbar
\dot{\vartheta}-U(r)\right] dt
\end{equation}
which we approximate by
\begin{equation}
S_{j}=\frac{M}{2\epsilon }
\left\{(\Delta r_{j})^{2} + 2\sigma ^{2}r_{j}r_{j-1}
\left[1 - \cos(\Delta \vartheta_{j})\right]\right\} - \xi \hbar
\Delta \vartheta_{j} - U_{j}\epsilon~,
\end{equation}
where $\Delta r_{j}=r_{j} - r_{j-1}$, $\Delta \vartheta_{j} = \vartheta_{j}
-\vartheta_{j-1}$, and $U_{j}=U(r_{j})$.
It is tempting to approximate $(d \vartheta)^{2}$ by $(\Delta
\vartheta)^{2}$. In path integration, however, $(\Delta
\vartheta)^{4}/\epsilon $ cannot be ignored because $(\Delta
\vartheta)^{2} \sim \epsilon $. Hence it is appropriate to replace
$(d\vartheta)^{2}$ by $1 - \cos(\Delta \vartheta)$ even though unimportant
higher order terms are included (see \cite{IKG}). Since it is sufficient
for a short time action to consider the contributions up to first order
in $\epsilon $, we further employ the approximate relation for small
$\epsilon $ \cite{IS},
\begin{equation}
\cos \left( \Delta \vartheta\right) + a\epsilon \Delta \vartheta \sim
\cos \left( \Delta \vartheta - a\epsilon \right)
+\frac{1}{2}a^{2}\epsilon ^{2}
\end{equation}
to write the short time action multiplied by $(i/\hbar)$ as
\begin{eqnarray}
\frac{i}{\hbar}S_{j}&=&\frac{iM}{2\hbar\epsilon }\left(r_{j}^{2} + r_{j-1}^{2}\right)
+(\sigma ^{-2} - 1)\frac{M\sigma ^{2}r_{j}r_{j-1}}{i\hbar \epsilon }
\nonumber \\
&&+ \frac{M\sigma ^{2}r_{j}r_{j-1}}{i\hbar \epsilon } \cos \left( \Delta
\vartheta_{j}-\frac{\xi \hbar \epsilon }{M\sigma ^{2}r_{j}r_{j-
1}}\right) + \frac{(4\xi ^{2} + \kappa )\hbar \epsilon}{8Mi\sigma
^{2}r_{j}r_{j-1}}-\frac{iM\omega ^{2}\epsilon }{4\hbar}(r_{j}^{2} +
r_{j-1}^{2}). \label{sact}
\end{eqnarray}
We use this short time action (multiplied by $i/\hbar$) for evaluating
the angular path integral. However, before starting the angular
integration, let us determine the amplitude $A_{j}$ by considering the
limit $\epsilon \rightarrow 0$ of the short time action (\ref{sact}),
\begin{eqnarray}
\lim_{\epsilon  \rightarrow 0}A_{j}e^{iS_{j}/\hbar} &= &
\lim_{\epsilon  \rightarrow 0}A_{j}\exp\left\{\frac{iM}{2\hbar\epsilon }
\left(\Delta r_{j}\right)^{2}\right\}\,
\exp\left\{\frac{iM\sigma ^{2}r_{j}^{2}}{2\hbar \epsilon }
\left(\Delta \vartheta_{j}\right)^{2}\right\} \nonumber \\
& = & A_{j}\frac{2\pi i\hbar \epsilon }{M\sigma r_{j}}
\delta (\Delta r_{j})\,\delta (\Delta \vartheta_{j}).
\end{eqnarray}
Now let the areal element be given by
\begin{equation}
d^{2}{\bf r}_{j}=r_{j}\,dr_{j}\,d\vartheta _{j}. \label{area1}
\end{equation}
Then the amplitude meeting the condition (\ref{norm2}) must be of the
form,
\begin{equation}
A_{j}= \frac{M\sigma }{2\pi i\hbar\epsilon}.
\label{amp1}
\end{equation}
Alternatively, if
\begin{equation}
d^{2}{\bf r}_{j} = \sigma r_{j}\,dr_{j}\,d\vartheta_{j}, \label{area2}
\end{equation}
then the corresponding amplitude is
\begin{equation}
A_{j}= \frac{M}{2\pi i\hbar\epsilon}. \label{amp2}
\end{equation}
Since path integration with either combination yields the same
result, we employ the first choice (\ref{area1}) with (\ref{amp1}) for
our calculation.

\subsection{Asymptotic recombination}

For path integration in polar coordinates, separation of variables is
not straightforward. To separate the angular variable from the radial
function, we employ the asymptotic recombination technique (see
\cite{IKG}). The asymptotic form of the modified Bessel function
$I_{\nu }(z)$ for large $|z|$ (GR:8.451.5 in \cite{GR}) is
\begin{equation}
I_{\nu }(z) \sim \frac{e^{z}}{\sqrt{2\pi z}} \sum_{n=0}^{\infty }(-
1)^{n}\frac{(\nu , n)}{(2z)^{n}} + \frac{e^{-z-(\nu + (1/2))\pi
i}}{\sqrt{2\pi z}}\sum_{n=0}^{\infty }\frac{(\nu ,n)}{(2z)^{n}}
\end{equation}
where $-3\pi /2 < \,\arg \,z \,< \pi /2$, and
\[
(\nu , n)= \frac{\Gamma (\nu +n+\frac{1}{2})}{n!\,\Gamma (\nu -n
+ \frac{1}{2})}, ~~~~~~~~(\nu , 0)=1.
\]
The asymptotic recombination technique is based on the conjecture that
the one-term asymptotic form,
\begin{equation}
I_{\nu }\left( z\right) \sim \frac{1}{\sqrt{2\pi z}}\exp \left[ z -
\frac{1}{2z}\left(\nu ^{2}-\frac{1}{4}\right) \right],  \label{EG}
\end{equation}
is valid for sufficiently large $|z|$ and for $-\pi /2 < \arg
\,z\, <\pi /2$ as relevant in path integration \cite{EG,LI}.
With the help of the one-term form which we refer to as the
Edwards-Gulyaev asymptotic formula, we can derive the following asymptotic
relation for large $|z|$ \cite{IKG},
\begin{equation}
I_{\nu }(az)\,e^{bz}\,e^{-c/z} \sim
\sqrt{\frac{a+b}{a}}\,I_{\mu }\left[(a+b)z\right], \label{asy1}
\end{equation}
where $a > 0$, $b \geq 0$, $8a(a+b)c> b - 4(a+b)\nu ^{2}$ and
\begin{equation}
\mu = \left[\frac{a+b}{a}\nu ^{2} - \frac{b}{4a} + 2(a+b)c \right]^{1/2}.
\end{equation}
Making use of this asymptotic relation together with the Jacobi-Anger
expansion formula
\begin{equation}
e^{z\cos \vartheta} =\sum\limits_{m=-\infty }^{\infty
}e^{im\vartheta}I_{m}(z)~,
\end{equation}
we obtain another asymptotic relation for large $z$,
\begin{equation}
\exp \left\{bz + z\cos \left[\Delta \vartheta + i\frac{d}{z}\right] -
\frac{(d^{2}+2f)}{2z}\right\} \sim \sqrt{1+b}\sum\limits_{m=-\infty
}^{\infty }e^{im\,\Delta \vartheta}\, I_{\mu }\left[(1+b)z\right]
\label{asy2}
\end{equation}
with
\begin{equation}
\mu =\left[(1+b)\{(m + d)^{2} +2f\}- b/4\right]^{1/2}.
\end{equation}
In the above we have let $a=1$ and $c=(d^{2} + 2f + 2md)/2$.

\subsection{Angular integration}

Utilizing (\ref{asy2}) with $b=\sigma ^{-2} -1$, $z=M\sigma
^{2}r_{j}r_{j-1}/(i\hbar \epsilon )$, $d=\xi =\alpha - \beta k$ and
$f=\kappa /2$,
we separate variables of the short time propagator (\ref{spro}) as
\begin{equation}
K^{(k)}\left( {\bf r}_{j},{\bf r}_{j-1};\epsilon \right)
=A_{j}\exp\left(\frac{i}{\hbar}S_{j}\right)=
\frac{1}{2\pi }\sum\limits_{m_{j}=-\infty }^{\infty }e^{im_{j}\left(
\vartheta_{j}-\vartheta_{j-1}\right) }R_{m_{j}}\left( r_{j},r_{j-
1};\epsilon \right)  \label{expS}
\end{equation}
with the short time radial propagator,
\begin{equation}
R_{m_{j}}\left( r_{j},r_{j-1};\epsilon \right) = \frac{M}{i\hbar
\epsilon} \exp \left[ \frac{iM}{2\hbar \epsilon }\left( r_{j}^{2}+r_{j-
1}^{2}\right) -\frac{iM\omega ^{2}\epsilon }{4\hbar }(r_{j}^{2}+r_{j-
1}^{2})\right] I_{\mu (m_{j})}\left( \frac{Mr_{j}r_{j-1}}{i\hbar
\epsilon }\right) \label{shtR}
\end{equation}
where
\begin{equation}
\mu (m_{j})=\frac{1}{2\sigma }
\left[4(m_{j} + \xi )^{2} + \sigma^{2}-1 + \kappa \right]^{1/2}. \label{mu}
\end{equation}
The prefactor $\sqrt{1+b}$ appeared on the right hand side of the
formula (\ref{asy2}) results in a multiple factor $\sigma ^{-1}$ before
the summation of (\ref{expS}), which will cancel out the factor $\sigma
$ appearing in the amplitude (\ref{amp1}). Thus the two-dimensional path
integral (\ref{disK}) is reduced to the form,
\begin{equation}
K^{(k)}\left({\bf r}^{\prime \prime }, {\bf r} ^{\prime };\tau \right)
=\lim_{N\rightarrow \infty } \int
\prod\limits_{j=1}^{N}\left[\frac{1}{2\pi
}\sum_{m_{j}}\,e^{im_{j}(\vartheta_{j} - \vartheta _{j-
1})}\,R_{m_{j}}(r_{j}, r_{j-1}; \epsilon )\right]\,
\prod\limits_{j=1}^{N-1} r_{j}dr_{j}\,d\vartheta_{j}.
\label{2pi2}
\end{equation}
The angular integration can be done straightforwardly by using the
orthogonality relation,
\begin{equation}
\int_{0}^{2\pi }e^{i(m'-m)\vartheta}\,d\vartheta
=2\pi \delta _{m', m}.
\end{equation}
Namely,
\begin{equation}
\int\,\prod_{j=1}e^{im_{j}\Delta \vartheta_{j}}\,\prod_{j=1}^{N-1}d\vartheta
_{j}=(2\pi )^{N-1} \prod_{j=1}^{N -1}\delta _{m, m_{j}}\,
e^{im(\vartheta''-\vartheta')}.
\end{equation}
where $m=m_{N}$. After angular integration, we have $m_{j}=m$ for all
$j$, and arrive at the expression
for the full propagator with a fixed wave number $k$,
\begin{equation}
K^{(k)}\left({\bf r}^{\prime \prime }, {\bf r}^{\prime }; \tau \right)
= \frac{1}{2\pi }\sum_{m=-\infty }^{\infty }\exp
\left[ im\left( \vartheta^{\prime \prime }-\vartheta^{\prime }\right)\right]
R_{m}(r'', r'; \tau ) \label{part}
\end{equation}
where
\begin{equation}
R_{m}(r'', r'; \tau )=\lim_{N\rightarrow \infty }
\int \prod\limits_{j=1}^{N}R_{m}\left( r_{j},r_{j-1};\epsilon \right)
\prod_{j=1}^{N-1} r_{j}dr_{j}. \label{rad}
\end{equation}
The left hand side of (\ref{part}) is nothing but the partial
wave expansion in two dimensions, and the radial propagator (\ref{rad})
corresponds to the $m$-th partial wave propagator. However,
the last radial path integration (\ref{rad}) remains to be carried out.

\subsection{Radial path integration}

To perform the radial path integration explicitly, we first notice that
the short time radial propagator (\ref{shtR}) with $m=m_{j}$ for all $j$
is similar in form as that of the radial harmonic oscillator. Then we
rewrite (\ref{rad}) as
\begin{equation}
R_{m}(r_{j},r_{j-1};\epsilon )=\frac{M}{i\hbar \epsilon }\exp
\left[\frac{iM\omega }{2\hbar} (r_{j}^{2}+r_{j-1}^{2})\frac{1}{\omega
\epsilon }\left(1-\frac{1}{2}\omega ^{2}\epsilon ^{2}\right)\right] \,
I_{\mu(m)}\left( \frac{Mr_{j}r_{j-1}}{i\hbar \epsilon }\right).
\label{shtR3}
\end{equation}
We further simplify (\ref{shtR3}) by letting $\eta _{j}=(M\omega
/2\hbar )r_{j}^{2}$ and $\varphi =\arcsin (\omega \epsilon )$ as
\begin{equation}
R_{m}(r_{j},r_{j-1};\epsilon )= \frac{M\omega }{i\hbar \sin \varphi }
\exp \left[i(\eta _{j} + \eta _{j-1}) \cot \varphi \right] \,I_{\mu
}\left(-2i\sqrt{\eta _{j}\eta _{j-1}}\,\csc \varphi \right).
\label{shtR4}
\end{equation}
Here we have used the approximation,
\begin{equation}
\cos \varphi _{j} = \cos[\arcsin (\omega \epsilon)] \sim  1 -
\frac{1}{2}\omega ^{2}\epsilon ^{2}.
\end{equation}
At this point, for convenience, we introduce a two point function,
referred to as the $\upsilon $-function \cite{IKG,IJ}, by
\begin{equation}
\upsilon _{\mu }\left( \eta ,\eta ^{\prime };\varphi \right) =-i\csc
\varphi \exp \left[ i\left( \eta +\eta ^{\prime }\right) \cot \varphi
\right] I_{\mu }\left( -2i\sqrt{\eta \eta ^{\prime }}\csc \varphi
\right)
\end{equation}
satisfying the convolution relation
\begin{equation}
\int\nolimits_{0}^{\infty }\upsilon _{\mu }\left( \eta ^{\prime \prime
},\eta ;\varphi \right) \upsilon _{\mu }\left( \eta ,\eta ^{\prime };\varphi
\right) d\eta =\upsilon _{\mu }\left( \eta ^{\prime \prime },\eta ^{\prime
};2\varphi \right) \label{conv}
\end{equation}
which can be derived from Weber's formula (GR: 6.633.2 in \cite{GR})
as modified in \cite{PI},
\begin{equation}
\int_{0}^{\infty }\,\exp(i\alpha r^{2})\,I_{\mu }(-iar)\,I_{\mu }(-
ibr)\,r\, dr=\frac{i}{2\alpha }\,\exp\left[-\frac{i}{4\alpha }(a^{2} +
b^{2}) \right] \,I_{\mu }\left(-\frac{ab}{2\alpha }\right)
\end{equation}
valid for Re $\alpha > 0$ and Re $\mu > -1$.
Then the short time radial function can be expressed in terms of the
$\upsilon $-function as
\begin{equation}
R_{m}(r_{j},r_{j-1};\epsilon )=\frac{M\omega }{\hbar }\upsilon _{\mu
(m)}\left( \eta _{j},\eta _{j-1};\varphi \right).
\end{equation}
Now we are ready to perform the radial path integration.
Substitution of this into (\ref{rad}) gives
\begin{equation}
R_{m}\left( r^{\prime \prime },r^{\prime };\tau \right) =\frac{M\omega
}{\hbar}\lim_{N\rightarrow \infty }\int
\prod\limits_{j=1}^{N}\upsilon _{\mu (m)}\left( \eta _{j},\eta _{j-
1};\varphi \right) \prod\limits_{j=1}^{N-1}d\eta _{j}.
\end{equation}
The convolution property (\ref{conv}) of the $\upsilon$-function
enables us to reach
\begin{equation}
R_{m}\left( r^{\prime \prime },r^{\prime };\tau \right)
=\frac{M\omega }{\hbar }
\upsilon _{\mu(m)}\left( \eta ^{\prime \prime },\eta ^{\prime };
\omega \tau \right).
\end{equation}
In the above we have also used the property,
\begin{equation}
\lim_{N \rightarrow \infty } (N\varphi )= \lim_{N \rightarrow \infty }
(N\,\sin \varphi)= \lim_{N \rightarrow \infty } (\omega N\epsilon) =
\omega \tau .
\end{equation}
In terms of the modified Bessel function it is written as
\begin{equation}
R_{m}\left( r^{\prime \prime },r^{\prime };\tau \right) =\frac{M\omega }{
i\hbar \sin \omega \tau }\exp \left[ \frac{iM\omega }{2\hbar }\left( r^{\prime
2}+r^{\prime \prime 2}\right) \cot \omega \tau \right] I_{\mu (m)}\left(
\frac{M\omega r^{\prime }r^{\prime \prime }}{i\hbar \sin \omega \tau }\right)
\label{rad3}
\end{equation}
with
\begin{equation}
\mu (m)=\frac{1}{2\sigma }\sqrt{4(m + \alpha - \beta k)^{2} + \sigma ^{2} - 1 +
\kappa}. \label{ind}
\end{equation}
Notice that $\mu (m)$ is a real positive number if $1- \sigma ^{2} <
\kappa $. In this manner we have completed the radial path integration
for the partial propagator with $k$ fixed. It turns out that the radial
propagator we have obtained above is identical in form with that of the
radial harmonic oscillator. The characteristics of the dispiration, the
flux tube, the uniform magnetic field and the assumed potential, which
make the present system different from the simple harmonic oscillator,
are all taken into the index $\mu (m)$ of the modified Bessel function.

With the radial propagator (\ref{rad3}), the full propagator with $k$
fixed is obtained by
\begin{eqnarray}
\lefteqn{K^{(k)}\left(r'', \vartheta ''; r', \vartheta '; \tau \right)}
\nonumber \\
&&=\frac{M\omega }{ 2\pi i\hbar \sin \omega \tau }\exp \left[
\frac{iM\omega }{2\hbar }\left( r^{\prime 2}+r^{\prime \prime 2}\right)
\cot \omega \tau \right] \,\sum_{m=-\infty }^{\infty }\,e^{im(\vartheta
''-\vartheta ')}\,I_{\mu (m)}\left( \frac{M\omega r^{\prime }r^{\prime
\prime }}{i\hbar \sin \omega \tau }\right) \label{Kfin}
\end{eqnarray}
or by converting $\vartheta $ into $\theta = \vartheta + \bar{\omega} t$,
\begin{eqnarray}
\lefteqn{K^{(k)}\left(r'', \theta ''; r', \theta '; \tau \right)}
\nonumber \\
&&=\frac{M\omega }{
2\pi i\hbar \sin \omega \tau }\exp \left[\frac{iM\omega }{2\hbar }\,
\left( r^{\prime 2}+r^{\prime \prime 2}\right) \cot \omega \tau \right]
\,\sum_{m=-\infty }^{\infty }\,e^{im(\theta ''-\theta ' - \bar{\omega}   \tau
)}\,I_{\mu (m)}\left( \frac{M\omega r^{\prime }r^{\prime \prime
}}{i\hbar \sin \omega \tau }\right). \label{Kfin2}
\end{eqnarray}

\section{Winding number expansion}

In the angular path integration performed above, we have not explicitly
taken account of the topological structure of the background medium
${\cal M}$. Since ${\cal M}={\bf R}^3\backslash\{x=y=0\}\cong {\bf R}\times {\bf R}^+\times S^1$,
the paths connecting two points, say $a$ and $b$, in ${\cal M}$ can wind around the $z$-axis many times, and may be classified (into homotopy classes) by the fundamental group $\pi_1 ({\cal M})=\pi_1 (S^1)\cong {\bf Z}$ of ${\cal M}$. A set of all homotopically equivalent paths in ${\cal M}$ is now characterized by a single winding number $n\in{\bf Z}$. Therefore the propagator $K(b,a)$ may be given as a sum of subpropagators $\tilde{K}_{n}(b,a)$
for the paths with different winding numbers $n$,
\begin{equation}
K(b,a)=\sum_{n\in {\bf Z}}C_{n}\,\tilde{K}_{n}(b,a).\label{111}
\end{equation}
The paths in a same class may be deformed into one another, so that the transition amplitudes corresponding to these paths must share the same phase factor, On the other hand, the paths belonging to different classes may have different phases. Laidlaw and deWitt \cite{LdW}, and Schulman \cite{SchT} argue that the coefficients $C_{n}$ are the one-dimensional unitary representations of the fundamental group $\pi_1 ({\cal M})\cong {\bf Z}$. From eq.\ (\ref{111}) it is apparent that $C_{n+m}=C_{m}C_{n}$. If no degeneracies and no internal degrees of freedom are assumed, $C_{n+1}=e^{i\delta}C_{n}$. Then the coefficients are given by the one-dimensional representations of the fundamental group $\pi_1 ({\cal M})$ or the additive group ${\bf Z}$, namely $C_n=e^{in\delta}$.

It has also been pointed out \cite{AI} that the angular momentum
representation and the winding number representation are complementary
to each other via Poisson's sum formula,
\begin{equation}
\sum_{n\in {\bf Z}}e^{2\pi in \xi }=
\sum_{m\in {\bf Z}}\delta (\xi - m).  \label{Pois}
\end{equation}
This means that the propagator $K^{(k)}({\bf r}'', {\bf r}'; \tau )$ we
have obtained in the preceding sections, as is given in the angular
momentum representation, cannot be regarded as a subpropagator carrying
a single winding number. Similarly, any partial propagator with a fixed
angular momentum quantum number cannot be decomposed to a full set of
subpropagators with winding numbers. The propagator obtained for the
dispiration field must be understood as a full bound state propagator
that contains contributions from all homotopically possible paths (see
\cite{IS,GI} for detail). Thus we look for a winding number representation of
$K^{(k)}({\bf r}'', {\bf r}'; \tau )$ in (\ref{Kfin}) via Poisson's
formula (\ref{Pois}).

Let us rewrite (\ref{part}) as
\begin{equation}
K^{(k)}({\bf r}'', {\bf r}'; \tau )=\frac{1}{2\pi }\int \sum_{m\in {\bf
Z}} \delta (\alpha ' +  \lambda - m)\,e^{i(\alpha'+\lambda) (\vartheta''-
\vartheta')}\,R_{\alpha '+ \lambda }(r'',r';\tau ) \,d\lambda  \label{Kfin3}
\end{equation}
with the radial propagator $R_{m}(r'',r';\tau )$ given by (\ref{rad3}).
Then we utilize Poisson's formula (\ref{Pois}) to convert (\ref{Kfin3})
into the winding number representation
\begin{equation}
K^{(k)}({\bf r}'', {\bf r}'; \tau )=\sum_{n\in {\bf Z}}
C_{n}\tilde{K}_{n}({\bf r}'',{\bf r}';\tau ).
\end{equation}
where
\begin{equation}
C_{n}= e^{i 2\pi n\alpha '} \label{coeff}
\end{equation}
and
\begin{eqnarray}
\lefteqn{\tilde{K}_{n}({\bf r}'',{\bf r}';\tau )} \nonumber \\
&& = \frac{M\omega e^{i\alpha '(\vartheta''-\vartheta')}}{2\pi
i\hbar\,\sin \omega \tau}\,
\exp \left[ \frac{iM\omega }{2\hbar }\left( r^{\prime
2}+r^{\prime \prime 2}\right) \cot \omega \tau \right]
\int_{-\infty
}^{\infty }\,e^{i\lambda (\vartheta''-\vartheta' - 2\pi n)}\,I_{\mu (\alpha '
+\lambda )}\left(
\frac{M\omega r^{\prime }r^{\prime \prime }}{i\hbar \sin \omega \tau
}\right) d\lambda.
\label{Kn}
\end{eqnarray}
In the coefficients $C_{n}$ of (\ref{coeff}) $\alpha '$ can chosen to
be an arbitrary real number. In particular, choosing $\alpha' =-\alpha$ one observes that the magnetic flux only appears as a pure phase factor $\exp\{-i\alpha(\vartheta'' - \vartheta '+2\pi n)\}$ in (\ref{Kn}). The choice $\alpha'=\beta k -\alpha$ furthermore shows that the screw dislocation has a similar effect, that is, it only appears as a phase $\exp\{-i(\alpha-k\beta)(\vartheta'' - \vartheta '+2\pi n)\}$ in (\ref{Kn}).
Evidently both belong to $U(1)$, the
one-dimensional unitary representation of $\pi _{1}({\cal M})$. The
result is consistent with the Laidlaw-deWitt-Schulman theorem albeit the
unitary factor is not unique.

\section{Energy spectrum and wave functions}

All the information concerning the energy spectrum and associated
radial wave functions for the bound states in two dimensions is
contained in the radial propagator (\ref{rad3}).
In order to extract such information out of (\ref{rad3}), we make use
of the Hille-Hardy formula (GR: 8.976.1 in \cite{GR}),
\begin{equation}
I_{\mu }\left(\frac{2\sqrt{xyz}}{1-z}\right)\frac{(xyz)^{-\mu /2}}{1-
z}\,\exp\left[-\frac{1}{2}(x+y)\frac{1+z}{1-z}\right]=
\exp\left[-\frac{x+y}{2}\right]\sum_{n=0}^{\infty
}\frac{n!\,z^{n}}{\Gamma (n + \mu +1)}\,L^{(\mu )}_{n}(x) \,L^{(\mu )}_{n}(y)
\label{iden2}
\end{equation}
where $L^{(\mu )}_{n}(x) $ is the Laguerre polynomial related to the
confluent hypergeometric function as
\begin{equation}
L^{(\mu )}_{n}(x) = \frac{\Gamma (n + \mu  + 1)}{\Gamma (\mu +1)\,n!}
\,F(-n, \mu +1 ; x).
\end{equation}
Notice that the confluent hypergeometric function is in general an
infinite series,
\begin{equation}
F(a, b; x)=\frac{\Gamma (b)}{\Gamma (a)}\,\sum_{s=0}^{\infty }
\frac{\Gamma (a + s )}{\Gamma (b + n)\,s!}\,x^{s}
\end{equation}
which converges only for $|x|< 1$ and becomes a polynomial for any $x$
when $a=0, -1, -2, ...$. Letting $x=(M\omega /\hbar)r'\,^{2}$,
$y=(M\omega /\hbar)r''\,^{2}$, and $z=e^{-2i\omega \tau }$ in
(\ref{iden2}),
we write (\ref{rad3}) as
\begin{eqnarray}
R_{m}\left( r'', r'; \tau \right) &=&\frac{2M\omega }{
\hbar}\left(\frac{M\omega }{\hbar}r'r''\right)^{\mu}
\exp \left[-\frac{M\omega }{2\hbar }\left( r'\,^{2} +
r''\,^{2}\right)\right]\, \nonumber \\
& & \times \sum_{n=0}^{\infty }\frac{n!\,e^{-
i\tau \omega (2n + \mu +1)}}{\Gamma (n+\mu  +1)}
\,L^{(\mu  )}_{n}\left(\frac{M\omega }{\hbar}r'\,^{2}\right)
\,L^{(\mu  )}_{n}\left(\frac{M\omega }{\hbar}r''\,^{2}\right). \label{rad4}
\end{eqnarray}
With this radial propagator the $k$-propagator (\ref{Kfin2})
can be cast into the form
\begin{equation}
K^{(k)}(r'',\theta '';r',\theta '; \tau )=\sum_{m=-\infty }^{\infty
}\sum_{n=0}^{\infty }\psi _{mn}(r'', \theta '')\psi^{\ast}
_{mn}(r', \theta ')\,e^{-i\tau \tilde{E}_{mn}/\hbar}, \label{k-pro}
\end{equation}
where
\begin{equation}
\tilde{E}_{mn}=\hbar \omega (2n + \mu (m) + 1) + m\hbar \bar{\omega }
\label{spec4}
\end{equation}
and
\begin{equation}
\psi _{mn}(r, \theta )=\sqrt{\frac{M\omega }{\pi\hbar}}
\sqrt{\frac{n!}{\Gamma (n+\mu +1)}}\left(\frac{M\omega }{\hbar}
r^{2}\right)^{\mu /2 }e^{-(M\omega/2\hbar ) r^{2}}\,L_{n}^{(\mu
)}\left(\frac{M\omega }{\hbar}\,r^{2}\right) e^{im\theta}. \label{psi1}
\end{equation}

Substitution of (\ref{mu}) into (\ref{spec4}) results in the
energy spectrum for the particle bound in two dimensions
\begin{equation}
\tilde{E}_{mn}=\hbar \omega \left\{2n + 1 + \frac{1}{2\sigma }\sqrt{4(m
+\alpha -\beta  k)^{2} + \sigma ^{2} - 1 + \kappa} \right\}+m\hbar\bar{\omega}
\label{spec5}
\end{equation}
where $n=0, 1, 2, \dots$, $m = 0, \pm 1, \pm 2, \dots$.
Adding the continuous spectrum (\ref{zspec}) as well as the ignored $V_0$ yields the full spectrum
of the system,
\begin{equation}
E_{mnk} = \tilde{E}_{mn} + \frac{\hbar^{2}k^{2}}{2M}+(\beta k - \alpha)\hbar\bar{\omega}.
\end{equation}
%\end{document}

\section{Concluding Remarks}

In concluding the present paper, we would like to make some remarks on the
discrete energy spectrum (\ref{spec5}) for the two-dimensional motion
around the dispiration. First we examine special cases.

(i) {\it The Landau levels}: ~The presence of the uniform
magnetic field is unimportant for the study on the dispiration. However,
if there are no dislocation, no disclination, no magnetic tube, no external
short-ranged repulsive and long-ranged attractive potential but the
uniform magnetic field, that is, if $\alpha = \beta = \kappa =0$,
$\sigma =1$ and $\omega _{0}=0$, then, as is expected, we have the
Landau levels,
\begin{equation}
E_{\bar{n},k} = 2\hbar \omega _{_{L}}\left(\bar{n} + \frac{1}{2}\right)
+ \frac{\hbar^{2}k^{2}}{2M}
\end{equation}
where $\omega _{_{L}}= eB/(2M)$ and $\bar{n} = n + (|m|+m)/2 = 0, 1, 2, \cdots$.

(ii) {\it The screw dislocation spectrum}: ~In the absence of the
uniform magnetic field, the disclination and the inverse-square
potential, i.e., in the case of $\bar{\omega }=0$, $\sigma =1$ (i.e.
$\kappa =0$), the discrete energy spectrum for the two-dimensional motion
(with fixed $k$) becomes
\begin{equation}
\tilde{E}_{mn} = \hbar \omega_{0} \left(2n + 1 +
|m + \alpha - \beta k|\right), \label{specii}
\end{equation}
which shows that the effect of the Burgers vector $b=2\pi \beta $ is
practically identical to that of the magnetic tube (the Aharonov-Bohm
effect) as pointed out in ref.\ \cite{Kawa}. In comparison with Wilczek's
anyon model \cite{Arov}, the screw dislocation plays a similar role of an
anyon by generating a fractional spin. From the geometrical point of
view, as we have seen earlier, the line dislocation causes torsion. This
spectrum explicitly shows that a source of torsion generates a spin
effect \cite{Heh,Ino}. In the limiting case of the vanishing flux tube and
the diminishing Burgers vector, we have simply the harmonic oscillator
spectrum,
\begin{equation}
E_{\bar{n}} = \hbar \omega_{0} (\bar{n} + 1),
\end{equation}
where $\bar{n}= 2n + |m| = 0, 1, 2, \cdots$.  The harmonic oscillator
potential is important in the present dislocation spectrum, without
which two-dimensional bound states cannot be formed in the vicinity of
the defect. In the calculation of the partition function for an anyon
gas, the oscillator potential has played a role of the regulator for
taming divergences \cite{Arov,IZ}.

(iii) {\it The wedge disclination spectrum}: ~If $\alpha =0$, $\beta =0$,
and $\bar{\omega }=0$, then the two-dimensional discrete spectrum takes
the form,
\begin{equation}
\tilde{E}_{mn}= \hbar \omega_{0} \left\{2n + 1 +
\frac{1}{2\sigma }\sqrt{4m^{2} + \sigma ^{2} -1 +
\kappa } \right\}, \label{speciii}
\end{equation}
which belongs to a particle bound near the disclination by the harmonic
oscillator potential plus a repulsive inverse-square potential. From
(\ref{si-be}) it is clear that $\sigma =1$ implies the vanishing deficit
angle $\gamma =0$. This corresponds to the absence of disclination. If
$0 < \sigma < 1$, then $2\pi  > \gamma >0 $, that is, the medium carries
a positive curvature at the center of disclination. In comparison with
the assumed square inverse repulsive potential term with $\kappa $
inside the square root of the spectrum, we see that the negative term
with $\sigma ^{2} -1$ represents the effect of an attractive force; we
may argue that the disclination effectively generates a short-ranged
attractive force around it. If $\sigma > 1$, the medium is negatively
curved, and a repulsive force is created around the center of
disclination (the saddle point).

The Schr\"odinger equations for the harmonic oscillator interacting
separately with dislocation and disclination have been solved by Furtado
and Moraes \cite{FM}. In the absence of the flux tube $\alpha =0$,
our dislocation spectrum (\ref{specii}) coincides with their result
obtain for the dislocation. However, our disclination spectrum
(\ref{speciii}) with $\kappa =0$ is not in agreement with theirs.

Finally, we examine for the case of disclination the difference between the
result from the Schr\"odinger equation and that of path integration.
Since some errors are involved in \cite{FM}, we present our own solution of
the Schr\"odinger equation which is slightly different from that in
\cite{FM}. The line element (\ref{metric}), if $\beta =0$, reads
\begin{equation}
ds^{2}=dr^{2} + \sigma ^{2}\,r^{2}\,d\theta ^{2} + dz^{2}.  \label{elem0}
\end{equation}
Although the Laplace-Beltrami operator can be used to write down the
Schr\"odinger equation as in \cite{FM}, we choose to let $\phi = \sigma
\theta $, and express (\ref{elem0}) as
\begin{equation}
ds^{2}=dr^{2} + r^{2}\,d\phi ^{2} + dz^{2},  \label{elem1}
\end{equation}
which is identical with the flat space line element except for $0 \leq
\phi <  2\pi \sigma.$ Then it is rather trivial to write down the
Schr\"odinger equation in terms of coordinates $(r, \phi , z)$; namely
\begin{equation}
-\frac{\hbar^{2}}{2M}\left\{\frac{1}{r}\frac{\partial }{\partial
r}\left(r \frac{\partial}{\partial r}\right) +
\frac{1}{r^{2}}\frac{\partial^{2}}{\partial \phi ^{2}} +
\frac{\partial^{2}}{\partial z^{2}}\right\}\psi + V(r)\psi =E\psi
\end{equation}
which can be easily solved for a two-dimensional harmonic oscillator
potential $V(r)=\frac{1}{2}M\omega ^{2}r^{2}$ with the periodic condition
$\psi (r, 2\pi \sigma +\phi, z)=\psi (r, \phi , z)$. The normalizable
solution is obtained in the form,
\begin{equation}
\psi (r, \phi , z)=N\,e^{ikz}e^{im\phi/\sigma }e^{-M\omega
r^{2}/2\hbar}\,r^{|m|/\sigma }\,F(-n_{r}, 1 + |m|/\sigma ; (M\omega
/\hbar)r^{2}), \label{S-psi}
\end{equation}
with the condition
\begin{equation}
\frac{1}{2}\left(1+\frac{|m|}{\sigma } - \frac{E}{\hbar\omega } +
\frac{k^{2}\hbar^{2}}{2M}\right)=-n_{r} ~~~~(n_{r}\in {\bf N}_{0}),
\end{equation}
which yields the energy spectrum,
\begin{equation}
E_{n_r m k}=\hbar\omega \left\{2n_{r} + 1 + \frac{|m|}{\sigma }\right\}+
\frac{k^{2}\hbar^{2}}{2M}.
\label{S-speciii}
\end{equation}
Here $m \in {\bf Z}$ as follows from the periodic condition. It is
apparent that the above spectrum differs from the path integral result
(\ref{speciii}). The term $\sigma ^{2}-1$ inside the square root of
(\ref{speciii}) lacks in (\ref{S-speciii}). Furthermore, the wave
functions are different. The wave functions (\ref{S-psi}) with $m\neq 0$
vanish at $r=0$, but the function with $m=0$ remains to be non-zero.
This is in contrast with the fact that the radial wave function
(\ref{psi1}) obtained by path integration vanishes at $r=0$ for
all values of $m$. In this treatment the nonvanishing singular curvature
at the disclination center $r=0$ plays no role. The Schr\"odinger
equation may have to be modified so as to accommodate the curvature
effect.

In general, the Schr\"odinger equation in curved space is written
in the form,
\begin{equation}
\left\{-\frac{\hbar^{2}}{2M}\Delta + V_{c}({\bf r}) + V({\bf r})
\right\}\psi ({\bf r}, t)=i\hbar\frac{\partial }{\partial t}\psi ({\bf
r}, t)      \label{Sch}
\end{equation}
where $\Delta $ is the Laplace-Beltrami operator, $V_{c}({\bf r})$ is
the potential due to the curvature effect and $V({\bf r})$ is any
external potential. Historically, Podolsky \cite{Pod} defined the
Schr\"odinger equation in curved space without the curvature term, i.e.,
$V_{c}({\bf r})=0$, but, as Schulman \cite{Schbook} puts it, there is no
reason, other than the prejudice for simplicity, to ignore the curvature term.
Comparing with Feynman's path integral, DeWitt \cite{BdW} has proposed
that the scalar curvature term is needed, that is, $V_{c}({\bf
r})=g\hbar^{2}R({\bf r})$ where $g$ is a constant. More recently,
however, Kleinert \cite{Kl3} has argued by using a quantum equivalence
principle that there is no need of the curvature term. On the other
hand, viewing that the motion in a two-dimensional curved space as a
constrained motion on the curved surface imbedded in a three-dimensional
Euclidean space, Jensen and Koppe \cite{JK}, da Costa \cite{daC} and
others have argued that the Schr\"odinger equation on a curved surface
carries in it an effect potential due to the Gaussian curvature $K({\bf
r})$ and the mean curvature $H({\bf r})$.

Our path integral calculation necessitates an effective potential of
the inverse square form which is due to neither the scalar curvature nor
the Gaussian curvature. In a forth coming paper \cite{next} it will be
shown that the path integration in a conical space with $K=0$ and
$H=\sqrt{1-\sigma ^{2}}/(2\sigma r)$ for $r\neq 0$ is compatible with the
Schr\"odinger equation modified with the mean curvature $H$ of the
conical surface as suggested in \cite{JK}.

\newpage
\textheight 240mm
\begin{figure}
\center{\bf Figures}\\[15mm]
  \includegraphics[width=200pt]{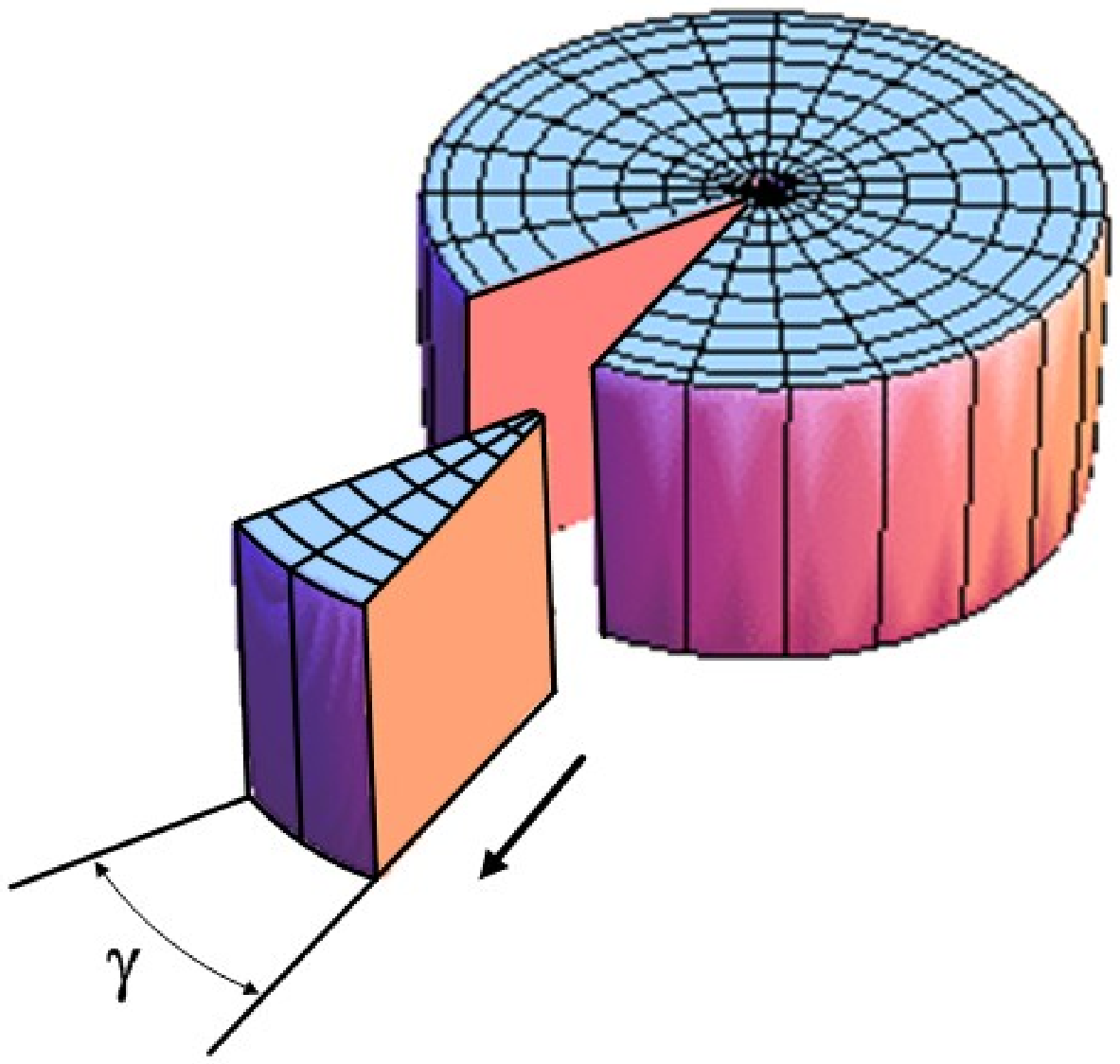}\\
  \caption{Make a thin cylindrical tube along the central line and remove a
  wedge of angle $\gamma >0 $ from the main body.}
  \label{Fig1}
~\\[5mm]
  \includegraphics[width=140pt]{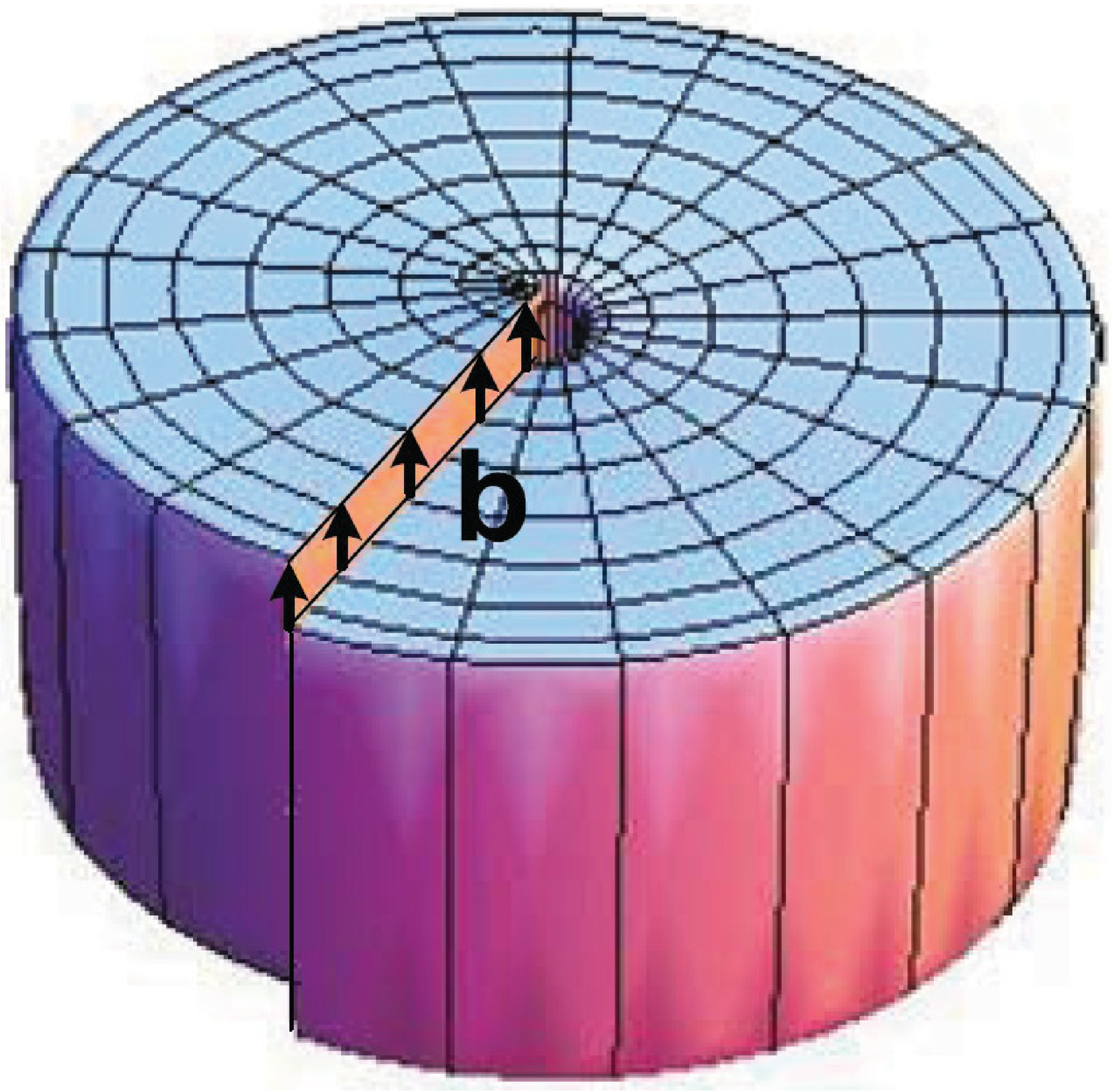}\\
  \caption{Close the open lips in such a way that a screw dislocation is
  created along the center line. The resulting medium is what we call the
  field of a dispiration.}
  \label{Fig2}
\end{figure}

\begin{thebibliography}{99}

\bibitem{RS} T.Y. Rebane and J.W. Steeds Phys. Rev. Lett. {\bf
75} 3716 (1995).
\bibitem{BST} R. Bausch, R. Schmitz and L.A. Turski, Phys. Rev. B {\bf 59}
13491 (1999).
\bibitem{A} S. Azevedo, J. Phys. A: Math. Gen. {\bf 34} 6081 (2001).
\bibitem{KF} M. Kleman and J. Friedel, Rev. Mod. Phys. {\bf 80} 61
(2008).
\bibitem{Kon1} K. Kondo, {\sl RAAG Memoirs of the Unifying Study of the
Basic Problems in Engineering and Physical Sciences by Means of Geometry},
vol 1 (Gakujutsu Bunken Fukyu-kai, Tokyo, 1952) and vol 2
(Gakujutsu Bunken Fukyu-kai, Tokyo, 1955).
\bibitem{Kon2} K. Kondo, in {\sl Proceedings of the 2nd Japan National
Congress for Applied Mechanics}, (Tokyo, 1952) p. 41; see also B.A.
Bilby, R. Bullough and E. Smith, Proc. Roy. Soc. A {\bf 231}
263 (1955).
\bibitem{Kro} E. Kr\"oner, in {\sl Physics of Defects}, Les Houches
Lectures, eds. B. Balian {\it et al}. (North Holland, Amsterdam, 1981).
\bibitem{KE} A. Kadi\'c and D.G.B. Edelen, {\sl A Gauge Theory of
Dislocations and Disclinations}, Lecture Notes in Physics 174
(Springer-Verlag, Berlin, 1983).
\bibitem{Kl1} H. Kleinert, {\sl Gauge Fields in Condenced Matter}
vol. II (World Scientific, Singapore, 1989).
\bibitem{PS} R.A. Puntigam and H.H. Soleng, Class. Quantum Grav.
{\bf 14} 1129 (1997), see also arXiv:gr-qc/9604057.
\bibitem{Kawa} K. Kawamura, Z. Phys. B {\bf 29} 101 (1978).
\bibitem{AB} Y. Aharonov and D. Bohm, Phys. Rev. {\bf 115} 485 (1959).
\bibitem{SchT} L.S. Schulman, J. Math. Phys. {\bf 12} 304 (1971).
\bibitem{Schbook} L.S. Schulman, {\sl Techniques and Applications of
Path Integration}, (Wiley, New York, 1981).
\bibitem{IS} A. Inomata and V.A. Singh, J. Math. Phys. {\bf 19} 2318
(1978).
\bibitem{GI} C.C. Gerry and A. Inomata, in {\sl Fundamental Questions in
Quantum Mechanics}, eds. L. Roth and A. Inomata (Gordon-Breach, New
York, 1986) p. 199.
\bibitem{Har} W.F. Harris, Philos. Mag. {\bf 22} 949 (1970).
\bibitem{T82}A.A. Tseytlin, Phys. Rev. D, {\bf 26}, 3327 (1982).
\bibitem{H95}F.W. Hehl, J.D. McCrea, E.W. Mielke and Y. Ne'emann, Phys. Rep.
{\bf 258}, 1 (1995).
\bibitem{I79}A. Inomata and M. Trinkala, Phys. Rev. D, {\bf 19},
1665 (1979).
%\bibitem{Iseri}See for example, D. Dietz and H. Iseri, {\sl Calculus and Differential Geometry:
%An Introduction to Curvature}, e-book available at http://faculty.mansfield.edu/hiseri/book-cdg.pdf .
\bibitem{Feyn}R.P. Feynman and A.R. Hibbs, {\sl Quantum Mechanics and Path
Integrals} (McGraw-Hill, New York, 1965).
\bibitem{EG} S.F. Edwards and Y.V. Gulyaev, Proc. Roy. Soc. London A
{\bf 279} 229 (1964).
\bibitem{PI}D. Peak and A. Inomata, J. Math. Phys. {\bf 10} 1422 (1969).
\bibitem{BJ}M. B\"ohm and G. Junker, J. Math. Phys. {\bf 28} 1978 (1987).
\bibitem{IKG} A. Inomata, H. Kuratsuji and C.C. Gerry, {\sl Path
Intgrals and Coherent States of SU(2) and SU(1,1)} (World Scientific,
Singapore, 1992).
\bibitem{GR} I.S. Gradshteyn and I.W. Ryzhik, {\sl Table of Integrals,
Series and Products} (Academic Press, New York, 1965).
\bibitem{LI} W. Langguth and A. Inomata, J. Math. Phys. {\bf 20} 499
(1979).
\bibitem{IJ}A. Inomata and G. Junker, in {\sl Noncompact Lie Groups
and Some of their Applications}, eds. E.A. Tanner and R. Wilson (Kluwer,
Dordrecht, 1994) p. 199.
\bibitem{LdW} M.G.G Laidlaw and C. Morette-DeWitt, Phys. Rev. D {\bf 3}
1375 (1971).
\bibitem{AI} A. Inomata, in {\sl New Techniques and Ideas in Quantum
Measurement Theory}, ed. D.M. Greenburger (Ann. New York Acad. Sci. {\bf 480},
(The New York Academy of Sciences, New York, 1986) p.217
\bibitem{Heh} F.W. Hehl and B.K. Data, J. Math. Phys., {\bf 12} 1334
(1971).
\bibitem{Ino} A. Inomata, Phys. Rev. D {\bf 18} 3552 (1978).
\bibitem{Arov} D. Arovas, in {\sl Geometric Phases in Phys},
eds. A. Schapere and F. Wilczek (Scientific, Singapore, 1989) p. 284.
\bibitem{IZ} A. Inomata and P.C. Zhu, in {\sl Path Integrals from meV to
MeV: Tutzing '92}, eds. H. Grabert, A. Inomata, L.Schulman and U. Weiss
(World Scientific, Singapore, 1993) p.136.
\bibitem{Wilc}F. Wilczek, Phys. Rev. Lett. {\bf 48} 1144 (1982);
Phys. Rev. Lett. {\bf 49} 957 (1982). See also F. Wilczek, {\sl
Fractional Statistics and Anyon Superconductivity} (World Scientific,
Singapore, 1990).
\bibitem{FM} C. Furtado and F. Moraes, J. Phys. A: Math. Gen.
{\bf 33} 5513 (2000).
\bibitem{Pod} B. Podolsky, Phys. Rev. {\bf 32} 812 (1928).
\bibitem{BdW} B.S. DeWitt, Rev. Mod. Phys. {\bf 29} 337 (1957).
\bibitem{Kl3} H. Kleinert, {\sl Path Integrals in Quantum Mechanics,
Statistics and Polymer Physics}, 2nd edn. (World Scientific, Singapore,
1995).
\bibitem{JK}H. Jensen and H. Koppe, Ann. Phys. (NY) {\bf 63} 589
(1971).
\bibitem{daC}R.C.T. da Costa, Phys. Rev. A {\bf 23} 1982 (1981).
\bibitem{next}A. Inomata and G. Junker, Phys.\ Lett.\ A {\bf 376} 305 (2012) and arXiv:1110.2279.
\end{thebibliography}
\end{document}